\documentclass[conference]{IEEEtran}
\IEEEoverridecommandlockouts
\usepackage{cite}
\usepackage{amsmath,amssymb,amsfonts}
\usepackage{algorithmic}
\usepackage{graphicx}
\usepackage{textcomp}
\usepackage{xcolor}
\usepackage{tabularx}
\usepackage{arydshln}
\usepackage{relsize}
\usepackage{balance}
\usepackage{boldline}
\def\BibTeX{{\rm B\kern-.05em{\sc i\kern-.025em b}\kern-.08em
    T\kern-.1667em\lower.7ex\hbox{E}\kern-.125emX}}
    
\newlength{\Oldarrayrulewidth}
\newcommand{\Cline}[2]{%
  \noalign{\global\setlength{\Oldarrayrulewidth}{\arrayrulewidth}}%
  \noalign{\global\setlength{\arrayrulewidth}{#1}}\cline{#2}%
  \noalign{\global\setlength{\arrayrulewidth}{\Oldarrayrulewidth}}}
  
\begin{document}

    \title{Indoor Localization Under Limited Measurements: A Cross-Environment Joint Semi-Supervised and Transfer Learning Approach\\
}

\author{  \IEEEauthorblockN{Mohamed I. AlHajri$^1$, Raed M. Shubair$^2$, Marwa Chafii$^3$}
  \IEEEauthorblockA{$^1$ Department of Electrical Engineering and Computer Science, Massachusetts Institute of Technology, USA\\
  $^2$ Department of Electrical and Computer Engineering, New York University (NYU) Abu Dhabi, UAE\\
  $^3$ ETIS UMR 8051, CY Paris Université, ENSEA, CNRS, F-95000, Cergy, France\\
                      Emails: malhajri@mit.edu, raed.shubair@nyu.edu, marwa.chafii@ensea.fr}
}

\maketitle

\begin{abstract}
The development of highly accurate deep learning methods for indoor localization is often hindered by the unavailability of sufficient data measurements in the desired environment to perform model training. To overcome the challenge of collecting costly measurements, this paper proposes a cross-environment approach that compensates for insufficient labelled measurements via a joint semi-supervised and transfer learning technique to transfer, in an appropriate manner, the model obtained from a  rich-data environment to the desired environment for which data is limited. This is achieved via a sequence of operations that exploit the similarity across environments to enhance unlabelled data model training of the desired environment. Numerical experiments demonstrate that the proposed cross-environment approach outperforms the conventional method, convolutional neural network (CNN), with a significant increase in localization accuracy, up to 43\%. Moreover, with only 40\% data measurements, the proposed cross-environment approach compensates for data inadequacy and replicates the localization accuracy of the conventional method, CNN, which uses 75\% data measurements.

\end{abstract} 

\section{Introduction}
The wide-scale proliferation of smart phones and wearable devices with wireless communication capabilities for 4G/5G has motivated the deployment of a wide range of location-based applications and services delivered to the user \cite{witrisal2016high,zafari2019survey}. Moreover, next-generation 6G and wireless networks based on millimeter-wave and THz frequencies are envisioned to be human-centric, which require accurate location estimation with centimeter-level accuracy \cite{dang2020should,sarieddeen2020next}. Consequently, estimating the location of the user or sensor node through indoor localization is becoming increasingly of paramount importance and a fundamental milestone for successful deployment of 5G and future 6G wireless network services and applications that fulfill the increasingly high requirements of data rate, energy efficiency, coverage and reliability.




Various localization techniques that employ the Wi-Fi infrastructure were reported in the literature such as indoor fingerprinting which constructs a database of indoor RF measurements in terms of received signal strength (RSS), channel transfer function (CTF), or frequency coherence function (FCF).  Several machine learning algorithms, such as $k$-nearest neighbour ($k$-NN), decision tree (DT), support vector machine (SVM), have been developed along with deep learning \cite{alhajri2019indoor,tian2018performance,alhajri2018classification,chen_novel_2020,njima2021convolutional,alhajri2020cascaded,njima2020deep}.  

Deep learning-based indoor localization faces the challenge of insufficient data available to perform training of the model \cite{njima2021indoor}. Obtaining sufficient measurements and labelling them is a tedious task that involves substantial human effort and resources to collect the costly measurements. To overcome such a challenge, this paper proposes a cross-environment approach which employs joint semi-supervised and transfer learning to leverage the similarity between environments and exploit the availability of unlabelled data, for enhancing model performance.

\section{Real Data from RF Measurements}

\subsection{Measurements Setup}
\begin{figure}[h]
       \centering
        \includegraphics[width=0.4\textwidth]{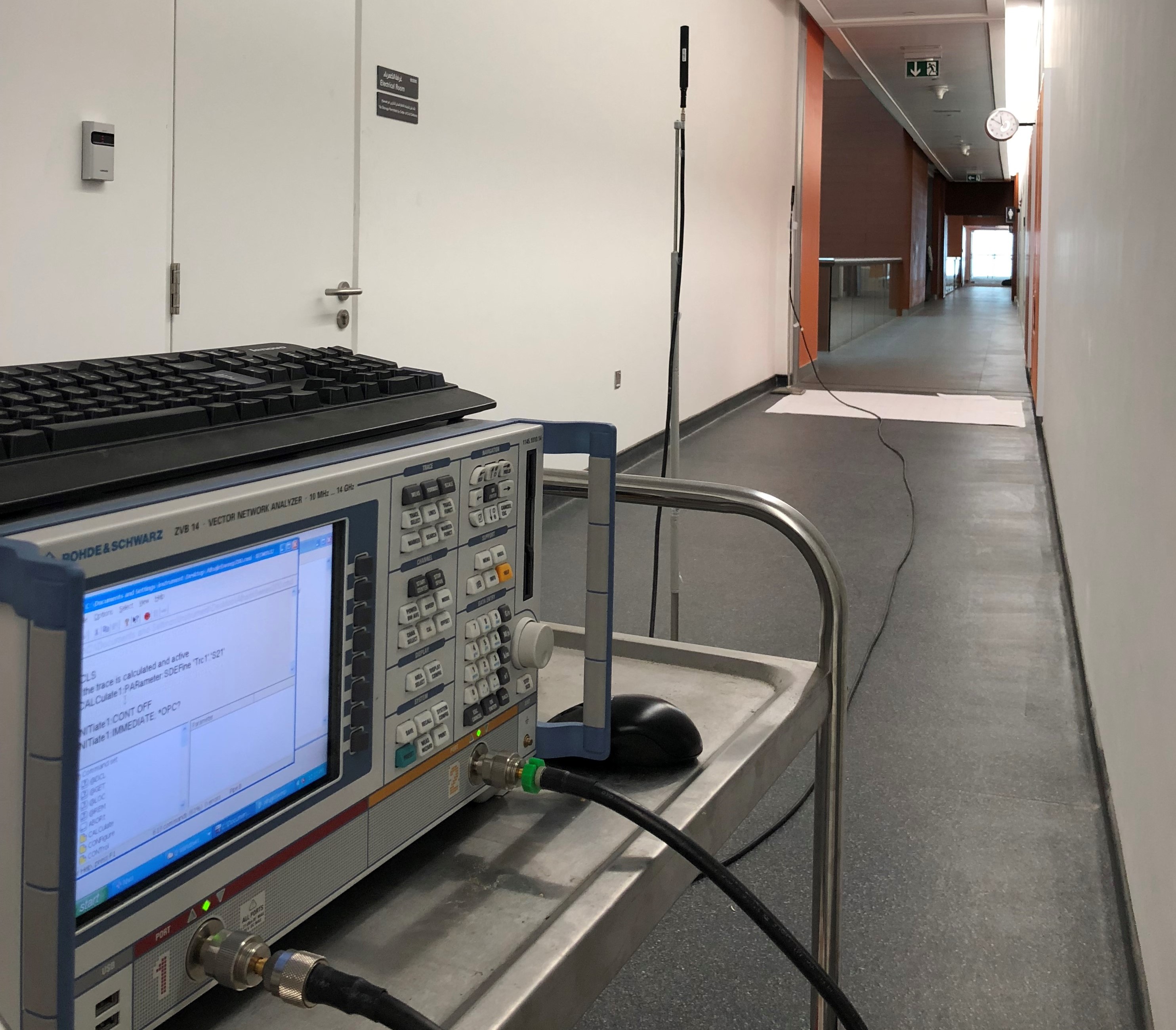} 
        \caption{Experimental measurement setup (left to right): vector network analyzer, transmit antennas, floor measurement grid, receive antenna. }
                \label{Map}
\end{figure}

Real data measurements were obtained using the setup in Fig.~\ref{Map}, which consisted of two omni-directional antennas (of height 1.5m each), low-loss RF cables, and a vector network analyzer (VNA) that measures $S_{21}$, which refers to CTF.  In fact, in the context of the physical environment, CTF refers to the measured complex value of the received signal, at position $(x,y)$, relative to the transmitted signal. Hence, CTF is an RF characteristic of the radio environment and shall be denoted by RF in this paper.  Under high signal-to-noise ratio (SNR) and for an operating frequency $f_o$, the received signal, is \cite{alhajri2018classification}:
\begin{equation}
\text{CTF}=\text{RF}(x,y)=\sum\limits^{L}_{l=1}a_l\exp{[-j(2\pi f_o\tau_1-\theta_l)]},
\label{1a}
\end{equation}%
where the three parameters $a_l$, $\tau_l$, and $\theta_l$ are respectively the location-dependent amplitude decay, time delay, and phase shift of the $l^\text{th}$ multipath component \cite{alhajri2018classification}.  These parameters are  \textit{spatial-dependent} in the sense that their values vary according to the location $(x,y)$ in the radio map of the environment, and $L$ is the total number of multipath components. 


\subsection{Frequency Points of RF Measurements}
A code script is executed for the VNA to run 10 consecutive sweeps at each frequency point within a 100MHz frequency band.  This process is repeated for 601 frequency points, for which the frequency separation is 166.667kHz such that $601 \times 166.667$kHz $=100$MHz. The measurements were carried out under a stationary scenario where we ensured no movements occur around the transmit and receive antennas. This ensured that the measured transmission coefficient will only account for the multipath components due to the physical environment without any additional distortion from moving objects.

\subsection{Location of Grid Points}
The experimental measurements setup was repeatedly assembled in four different locations within a typical university campus. These locations were chosen to represent the primary types of any typical real-world indoor environment which are: highly cluttered environment (laboratory), medium cluttered  environment (narrow corridor), low cluttered environment (lobby), and open space  environment (sports hall). For each type of the four indoor  environments, readings of the RF signal were taken at positions that are not close to each other but also not far apart; ensuring that small-scale variations can still be captured \cite{chen_novel_2020}. This has been achieved by arranging a square floor grid area in the measurements scene, and dividing it into uniform square cells, each having 12.5cm side length which is equivalent to one wavelength at frequency 2.4GHz being selected to examine the WiFi bands associated with IEEE 802.11g standard. This square-cell division of the floor grid results in a total of $14 \times 14=196$ cells. The receive antenna is moved across the floor grid and is positioned at the corners of the cells. At each cell corner, measurements on the VNA are obtained for 601 frequency points with 10 sweep readings each.

\subsection{Formation of Dataset}
The number of samples that comprise the dataset for each environment is the product of (196 grid positions $\times$ 601 discrete frequency points $\times$ 10 sweeps per frequency point).  

Once entirely constructed from measurements, the full dataset is divided into two portions:  75\% of the data samples are used for training the prediction algorithm, whereas the remaining 25\% data samples are used for testing and validation.

\subsection{Dataset Made Available as Open-Access}
We are pleased to share  the generated dataset with the academic community.  It has been made available online for further explorations and can be found at IEEE data port and is available online as open access \cite{ggh1-6j32-18}. The same dataset is also available online on University of California Irvine (UCI) Machine Learning Repository \cite{UCIML24GHZ}.    

\begin{figure*}[h]
    \centering
    \includegraphics[width = 0.90\textwidth]{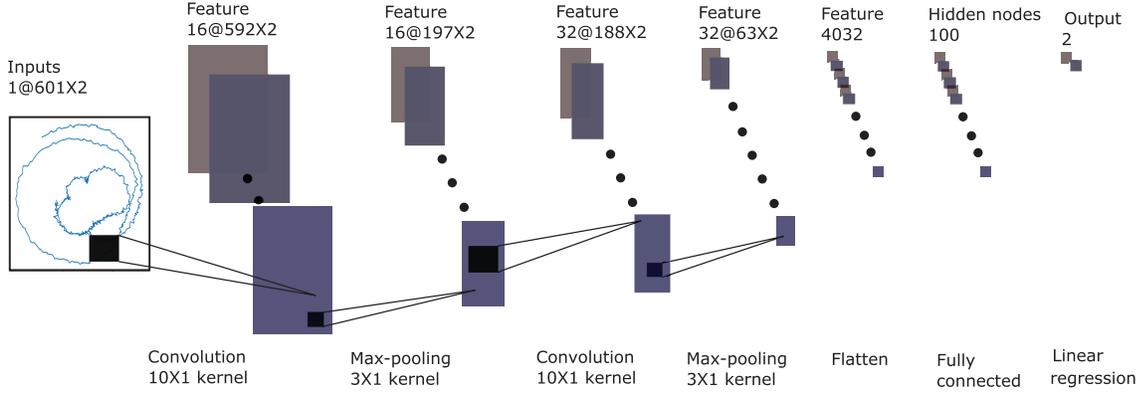}
    \caption{Architecture of CNN being the building block of the proposed cross-environment approach.}
    \label{fig:framework}
\end{figure*}

\section{Proposed Cross-Environment Approach}
\subsection{Using CNN as the Building Block}
Convolutional neural networks (CNNs), depicted in Fig.~\ref{fig:framework}, are known for their strong ability of learning spatial representations from raw data \cite{lecun2015deep}. This attribute makes CNN a favorable method for several applications. Moreover, CNN is employed in this paper to jointly optimize feature learning and inference, through an end-to-end trainable architecture of Fig.~\ref{fig:framework}.  


Let $\mathbf{X} \in \mathbb{R}^{\alpha \times \beta}$ be the input data to the convolution layer, where $\alpha=601$ is the number of frequency points, and $\beta =2$ to denote the real and imaginary values of CTF. The convolutional operation uses $P$ filters $\mathbf{K}_i \in \mathbb{R}^{m \times n}$, where in general $m \leq \alpha ~\text{and}~ n \leq \beta$, to sweep over the input matrix $\mathbf{X}$ from the top left to bottom right corner. One convolution operation is defined as: 
\begin{align}
    o_i(s,t) = \sigma \left(\sum_{b=1}^n \sum_{a=1}^m \tilde{\mathbf{X}}(s+a,t+b)\mathbf{K}_i(a,b) \right),
\end{align}
where $\tilde{\mathbf{X}} \in \mathbb{R}^{m \times n}$ is part of the input matrix $\mathbf{X}$, and $\sigma(.)$ is the activation function, adopted to be the rectified linear unit (ReLU) in this work. 
The output from the convolution operation is $\mathbf{O} \in \mathbb{R}^{P \times (\alpha - m + 1) \times (\beta-n + 1)} = \mathbb{R}^{16\times 592 \times 2}$. 

The function of the pooling layer is to reduce model parameters by reducing the number of features. This is achieved by selecting the dominant features which are rotationaly- and positionaly-invariant. In the CNN framework, we adopt  max-pooling, which returns the maximal value of its inputs. Given a max-pooling size of $d \times e$, the output of the max-pooling layer will have a dimension of $P \times \left \lfloor(\alpha - m + 1)/d \right \rfloor \times \left \lfloor(\beta-n + 1)/e\right \rfloor$, where $\left \lfloor \cdot \right \rfloor$ is the rounding down operation.

Stacking multiple layers allows for extracting more abstract features from raw data \cite{he2016deep}. For a stacked structure, the output of the previous layer is the input of the next layer, such that feature information is propagated through space. While the first layer tends to learn general features, the last layer tends to learn task-specific features \cite{yosinski2014transferable}. 
In our CNN model, we stack two convolution-pooling layers to learn high-level features from the raw data, which is shown in Fig. \ref{fig:framework}. 

The outputs of the convolutional network are  flattened into a column vector that is fed into a fully-connected layer used to insert non-linearity into features and seek more abstract representation. With $\boldsymbol{\mathrm{\mu}} \in \mathbb{R}^{4032 \times 1}$ being resulting vector after flattening, the output $\boldsymbol{\mathrm{\eta}} \in \mathbb{R}^{100 \times 1}$ of the fully-connected layer is given by:
\begin{align}
    \boldsymbol{\mathrm{\eta}}=\sigma \left ( \boldsymbol{\mathrm{W}}^{f}\boldsymbol{\mathrm{\mu}}  + \boldsymbol{\mathrm{\nu}}^{f} \right ),
\end{align}
where $\boldsymbol{\mathrm{W}}^{f} \in \mathbb{R}^{100 \times 4032}$ and $\boldsymbol{\mathrm{\nu}}^{f} \in \mathbb{R}^{100 \times 1}$ are respectively the weights and biases of the fully-connected layer. 

The output vector $\boldsymbol{\mathrm{\eta}}$ of the fully-connected layer is input to a linear regression layer that performs localization:
\begin{align}
    \mathbf{y} = \text{linear}(\boldsymbol{\mathrm{W}}^{l} \boldsymbol{\mathrm{\eta}} + \boldsymbol{\mathrm{\nu }}^{l}),
\end{align}
where $\boldsymbol{\mathrm{W}}^{l} \in \mathbb{R}^{2 \times 100}$ and $\boldsymbol{\mathrm{\nu }}^{l} \in \mathbb{R}^{2 \times 1}$ are the weights and biases of the linear regression layer. The output $\mathbf{y} \in \mathbb{R}^{2 \times 1}$ has two elements which correspond to the target position $(x,y)$.

Given the true labels and the model outputs, errors are calculated and backpropagated to update the model parameters via gradient-based optimization algorithms. We use the optimization algorithm of \emph{Adam}  \cite{kingma2014adam} to compute adaptive learning rates for different parameters. We also use \emph{dropout} \cite{srivastava2014dropout} after the fully-connected layer and batch normalization \cite{NIPS2018_7515} for each convolution layer to prevent over-fitting.


\subsection{Joint Semi-Supervised and Transfer Learning}
The proposed cross-environment joint semi-supervised and transfer learning approach is divided into six successive stages, as illustrated in Fig.~\ref{proposed}. We utilize data from two environments: (1) rich labelled measurement database referred to as \textit{source environment}, and (2) limited labelled measurement database referred to as \textit{target environment}. Datapoints in the source environment are denoted by $\mathbf{RF}_s^l \in \mathbb{R}^{601 \times 2}$ and labelled with respect to their associated position referred to by $\mathbf{C}_s \in \mathbb{R}^{2 \times 1}$ for simplicity. Each datapoint at the associated position is the CTF comprising 601 frequency points each of which is associated with a complex-valued received signal reading. On the other hand, datapoints in the target environment are both labelled $\mathbf{RF}_t^l \in \mathbb{R}^{601 \times 2}$ and unlabelled $\mathbf{RF}_t^u  \in \mathbb{R}^{601 \times 2}$. The labels of $\mathbf{RF}_t^l$ denote to the associated position referred to by $\mathbf{C}_t \in \mathbb{R}^{2 \times 1}$ for simplicity.


In the first stage of Fig.~\ref{proposed} which describes the proposed cross-environment approach, CNN is trained using the source environment labeled data samples $\mathbf{RF}_s^l$.  Accordingly, the loss function to be minimized is the following mean square error:

\begin{figure*}[h]
       \centering
        \includegraphics[width=0.40\textwidth,bb=157 35 449 749,angle=-90]{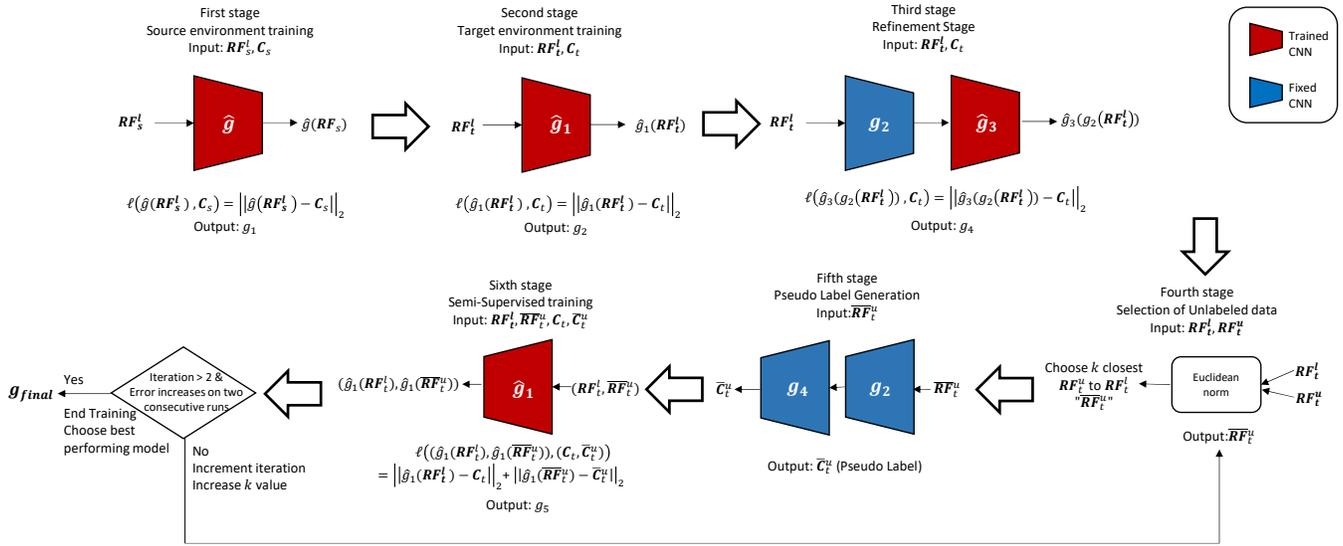}
        \caption{Proposed cross-environment joint semi-supervised and transfer learning approach.} 
                \label{proposed}
\end{figure*}

\begin{equation}
\ell(\hat{g}(\mathbf{RF}_s^l),\mathbf{C}_s) = ||\hat{g}(\mathbf{RF}_s^l)  - \mathbf{C}_s||_2,
\end{equation}
where $\hat{g}(.)$ is the trained CNN model initialized randomly, for which the input is ($\mathbf{RF}_s^l$) and output is $\hat{g}(\mathbf{RF}_s^l)$ representing the estimated position. The resulting CNN function is denoted by $g_1$, which has a very good localization performance since it is trained with the rich data of the source environment. 

In the second stage of Fig.~\ref{proposed} which describes the proposed cross-environment approach, the CNN model is trained by minimizing a loss function comprised of the mean squared error  between $\mathbf{C}_t$ and $\hat{g}_1(\mathbf{RF}_t^l)$:

\begin{equation}
\ell(\hat{g}_1(\mathbf{RF}_t^l),\mathbf{C}_t) = ||\hat{g}_1(\mathbf{RF}_t^l) - \mathbf{C}_t ||_2 
\end{equation}
where $\hat{g}_1(\mathbf{RF}_t^l)$ is the trained CNN model, initialized with $g_1$ (output of the first stage). The resulting CNN model of the second stage is denoted by $g_2$. 

The third stage of the proposed cross-environment approach shown in Fig. \ref{proposed} is a refinement stage that improves the estimation of the location, which is important for the generation of the pseudo labels of the unlabelled data.  In this stage, the DNN model is trained by minimizing a loss function comprised of the mean squared error between $\mathbf{C}_t$ and $\hat{g}_3(g_2(\mathbf{RF}_t^l)) \in \mathbb{R}^{2 \times 1}$:

\begin{equation}
\ell(\hat{g}_3(g_2(\mathbf{RF}_t^l)),\mathbf{C}_t) = ||\hat{g}_3(g_2(\mathbf{RF}_t^l)) - \mathbf{C}_t ||_2 
\end{equation}
where $\hat{g}_3$ is the trained DNN model, initialized with $g_1$ (output of the first stage). The resulting DNN model of the third stage is denoted by $g_4$. 

In the fourth stage of Fig.~\ref{proposed}, the Euclidean norm between unlabelled target data $\mathbf{RF}_t^u$ and labelled target data $\mathbf{RF}_t^l$ is calculated. The $k$ $\mathbf{RF}_t^u$ of the smallest Euclidean norm in comparison to the $\mathbf{RF}_t^l$ is selected for use in the upcoming stages. The selected subset of $\mathbf{RF}_t^u$ is denoted by $\overline{\mathbf{RF}}_t^u$. 

In the fifth stage of Fig.~\ref{proposed} which describes the proposed cross-environment approach, a pseudo label is generated for $\overline{\mathbf{RF}}_t^u$ using models $g_2$ and $g_4$ and will be denoted $\overline{\mathbf{C}}_t^u$. 

In the sixth stage  of Fig.~\ref{proposed} which describes the proposed cross-environment approach, $\mathbf{RF}_t^l$ and $\overline{\mathbf{RF}}_t^u$ will be used to train the CNN model by minimizing the loss function comprised of the sum of mean squared errors between $(\mathbf{C}_t,\overline{\mathbf{C}}_t^u)$ and the corresponding $(\hat{g}_1(\mathbf{RF}_t^l),\hat{g}_1(\overline{\mathbf{RF}}_t^u))$, respectively:
\begin{align}
&\ell((\hat{g}_1(\mathbf{RF}_t^l),\hat{g}_1(\overline{\mathbf{RF}}_t^u)),(\mathbf{C}_t,\overline{\mathbf{C}}_t^u)) \nonumber \\ &= ||\hat{g}_1(\mathbf{RF}_t^l) - \mathbf{C}_t ||_2 +||\hat{g}_1(\overline{\mathbf{RF}}_t^u) - \overline{\mathbf{C}}_t^u ||_2 
\end{align}
where the resulting CNN model of the sixth stage is denoted by $g_5$. The last three stages in Fig. \ref{proposed} will be repeated where the size of the unlabelled data $\mathbf{RF}_t^u$ will be increased. The final CNN, $g_{final}$, is the $g_5$ model that produces the least position estimation error (RMSE).


\section{Results and Discussion}

Localization accuracy is assessed in terms of  the error, RMSE, between estimated and labelled positions:
\begin{equation}
    \text{RMSE} = \left[\dfrac{1}{n}\sum_{i=1}^n (\overline{x}_{i} - x_i)^2 + (\overline{y}_{i} - y_i)^2\right]^{1/2}
\end{equation}

In the results to follow, conventional CNN method refers to applying CNN directly to the limited-data target environment.

\subsection{Degradation of Conventional CNN Due to Limited Data}
Table \ref{tab:my-table} summarizes the results of training a CNN in a conventional way using different sizes of the training data. It is evident that reducing the size of the training data degrades performance since the estimated position error (RMSE) increased. This is due to model overfitting and that the CNN is not able to generalize and predict new observed data. 

\begin{table}[h]
\centering
\caption{Reducing training data size degrades CNN performance: Estimated position error (RMSE) increases.}
\label{tab:my-table}
\renewcommand{\arraystretch}{1.3}
\scalebox{0.95}{
\begin{tabular}{c|c|c|c|c|}
\cline{2-5}
\multicolumn{1}{l|}{}                 & \multicolumn{4}{c|}{Size of Available Data for Training}                                                           \\ \cline{2-5}
                                      & {75\%} & {50\%} & {30\%} & {10\%} \\ \hline \Cline{1pt}{2-5}
\multicolumn{1}{|c|}{Lab}             & 8.46 cm         & 12.06 cm        & 13.91 cm       & 28.6 cm       \\ \hline
\multicolumn{1}{|c|}{Narrow Corridor} & 10.72 cm         & 11.90 cm        & 18.93 cm       & 37.5 cm       \\ \hline
\multicolumn{1}{|c|}{Lobby}           & 8.21 cm         & 10.03 cm        & 14.39 cm       & 34.07 cm       \\ \hline
\multicolumn{1}{|c|}{Sport Hall}      & 8.33 cm         & 11.44 cm        & 14.90 cm       & 33.54 cm       \\ \hline
\end{tabular}}
\end{table}

\begin{table*}[h]
\centering
\caption{Proposed cross-environment approach at low training data sizes: Joint semi-supervised and transfer learning \\ Gain in localization accuracy due to reduction in estimated position error (RMSE). }
\label{tab:my-table2}
\renewcommand{\arraystretch}{1.8}
\resizebox{\textwidth}{!}{%
\begin{tabular}{c|c|c|c|c|c|c|c|c|c|c|c|c|}
\cline{2-13}
\multicolumn{1}{l|}{}                 & \multicolumn{3}{c|}{Training Data Size: 15\%} & \multicolumn{3}{c|}{Training Data Size: 10\%} & \multicolumn{3}{c|}{Training Data Size: 5\%} & \multicolumn{3}{c|}{Training Data Size: 2.5\%} \\ \cline{2-13} 
\multicolumn{1}{l|}{}                 & Conventional  & Proposed  & Accuracy Gain     & Conventional  & Proposed  & Accuracy Gain     & Conventional  & Proposed  & Accuracy Gain    & Conventional   & Proposed  & Accuracy Gain     \\ \hline
\multicolumn{1}{|c|}{Lab}             & 22.88         & 13.75     & \textbf{39.90\%}  & 28.60          & 20.36     & \textbf{28.81\%}  & 42.92         & 26.76     & \textbf{37.65\%} & 44.4           & 31.54     & \textbf{28.96\%}  \\ \hline
\multicolumn{1}{|c|}{Narrow Corridor} & 30.33         & 21.82     & \textbf{28.06\%}  & 37.50          & 29.6      & \textbf{21.07\%}  & 58.5          & 42.00        & \textbf{28.21\%} & 58.44          & 51.94     & \textbf{11.12\%}  \\ \hline
\multicolumn{1}{|c|}{Lobby}           & 20.78         & 16.32     & \textbf{21.46\%}  & 34.07         & 19.46     & \textbf{42.88\%}  & 51.64         & 30.05     & \textbf{41.81\%} & 55.63          & 37.88     & \textbf{31.91\%}  \\ \hline
\multicolumn{1}{|c|}{Sports Hall}     & 28.5          & 17.33     & \textbf{39.19\%}  & 33.54         & 21.51     & \textbf{35.87\%}  & 47.63         & 30.70      & \textbf{35.54\%} & 44.60           & 38.47     & \textbf{13.74\%}  \\ \hline
\end{tabular}}
\end{table*}
\subsection{Proposed Approach Compensates for Limited Data}

\begin{figure}[h]
       \centering
        \includegraphics[width=0.5\textwidth]{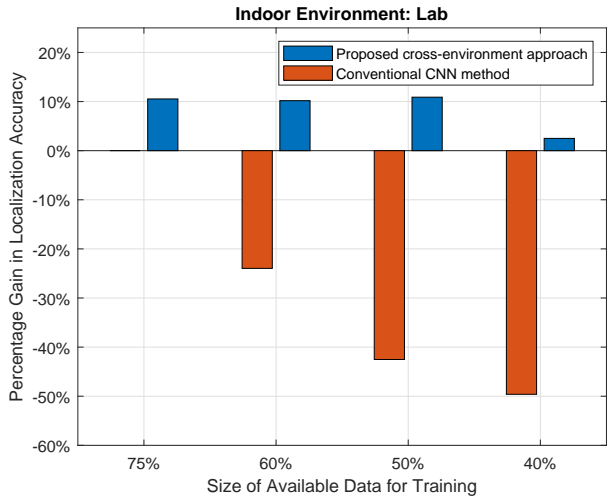}
        \caption{Increased accuracy achieved using proposed cross-environment approach compared to conventional CNN method, for different training data sizes. }
                \label{Comparision}
\end{figure}

Fig. \ref{Comparision} plots the percentage gain in localization accuracy (or equivalently percentage reduction in RMSE) versus the size of available data for training. We have obtained the results for the "Lab" being the target environment, where the source environment is the "Lobby".  Results of the proposed cross-environment approach are compared to the baseline conventional case: CNN trained on 75\% of the available data.  It is evident that the proposed cross-environment approach outperforms the conventional CNN method with significant performance gains.  While the proposed cross-environment approach  can still provide an accuracy gain at low data sizes (as low as 40\% of available data), the performance of conventional CNN method degrades severely, starting from 60\% training data size.  Furthermore, Fig. \ref{Comparision} demonstrates that, with only 40\% of the available data, the proposed cross-environment approach is  capable of compensating for the data deficit and replicates the localization accuracy of the conventional CNN method which is trained on 75\% of the available data.

Beside the aforementioned findings obtained for the Lab as a sample indoor environment, we expanded our numerical experiments to investigate all types of indoor environments with results shown in Table \ref{tab:my-table2}.  It is evident from  Table \ref{tab:my-table2} that the proposed cross-environment approach provides a significant increase in localization accuracy up to 43\%, compared to the conventional CNN method.  This is attributed to the joint semi-supervised and transfer learning approach which leverages the similarity between environments and exploits the availability of unlabelled data.

\section{Conclusion}
This paper tackled the challenge of developing highly accurate indoor localization deep learning methods under limited data measurements. The paper proposed a cross-environment approach that compensates for insufficient labelled measurements via a joint semi-supervised and transfer learning technique to transfer, in an appropriate manner, the model obtained from a  rich-data environment to the desired environment for which data is limited. This is achieved via a sequence of operations that exploit the similarity across environments to enhance unlabelled data model training of the desired environment. Numerical experiments demonstrated that the proposed cross-environment approach outperforms the conventional method, CNN, with a significant increase in localization accuracy, up to 43\%. Moreover, with only 40\% data measurements, the proposed cross-environment approach compensates for data inadequacy and replicates the localization accuracy of the conventional method, CNN, which uses 75\% data measurements.


\bibliographystyle{IEEEtran}
\bibliography{main}

\end{document}